# A CMOS Compatible Aluminum Scandium Nitride-based Ferroelectric Tunnel Junction Memristor

Xiwen Liu,[1] Jeffrey Zheng,[2] Dixiong Wang,[1] Pariasadat Musavigharavi,[1] Eric A. Stach,[2] Roy Olsson III[1] and Deep Jariwala[1, a)]

[1]*Department of Electrical and System Engineering, University of Pennsylvania, PA, 19104, USA*

[2] *Department of Material Science and Engineering, University of Pennsylvania, PA, 19104, USA*

**Abstract:** In this letter, we report a back-end-of-line (BEOL), complementary-metal–oxide–semiconductor (CMOS) compatible $Al_{0.64}Sc_{0.36}N$-based ferroelectric tunnel junction (FTJ) memristor. We demonstrate a memristor comprising a metal/insulator/ferroelectric/metal structure (Pt/native oxide/$Al_{0.64}Sc_{0.36}N$/Pt) that is compatible with BEOL temperatures (≤ 350 °C) grown on top of a 4-inch silicon wafer. The device shows self-selective behavior as a diode with > $10^5$ rectification ratio (for 5 V). It can suppress sneak currents without the need for additional access transistors or selectors. Furthermore, the memristor has an On/Off ratio of ~ 50,000 between low and high resistance states. The FTJ memristor has a stable programmed state during DC cycling and a retention time longer than 1,000 s at 300K. These results demonstrate that this system has significant potential as a future high-performance post-CMOS compatible non-volatile memory technology.

New device applications such as Internet of Things (IoT) devices, non-von Neumann computing architectures, and artificial intelligence (AI) computing algorithms are creating strong demand for high-density non-volatile memory (NVM) solutions with low power consumption. Among various emerging NVM technologies [1], ferroelectric random-access memories (FeRAM) are compelling due to their high access speed, high endurance, extremely low write energy and current, and good retention [2-5]. However, the incorporation of FeRAM into commercial scale semiconductor applications has been stalled at the 130-nm node [5]. There are three main challenges that have hindered rapid development of ferroelectric (FE) memories and has kept them from challenging classical charge-based memories and other NVM technologies: (1) the traditional one transistor–one ferroelectric capacitor (1T1C) structure undergoes destructive readout and has a large footprint; (2) traditional ferroelectrics are incompatible with BEOL, CMOS process; (3) as perovskite FE materials such as lead zirconium titanate (PZT) or barium titanate (BTO) are scaled to thinner layers their ferroelectric properties degrade.

a) Author to whom correspondence should be addressed. Electronic mail: dmj@seas.upenn.edu.

Various emerging technologies have been developed over the past decade to address these issues. The ferroelectric tunnel junction (FTJ) memristor, which utilizes a polarization-dependent tunneling current to perform resistive switching, is a promising alternate to a 1T1C cell. This is because it has the advantage of being a compact, two-terminal geometry device that uses a non-destructive read-out [4-5]. Recently, doped $HfO_2$ based FTJ memristors have attracted considerable attention, however, the high annealing temperatures ($\geq$ 400 °C) necessary for doped $HfO_2$ to attain ferroelectricity renders them unsuitable for CMOS BEOL process integration [4-5]. Further, FTJs based on $HfO_2$ have thus far shown a limited On/Off ratio, on the order of ~10 [4-5]. A high On/Off ratio is essential not only to enable low power, in-memory computing but also for maintaining a strong immunity to noise and variations when used in emerging applications such as multi-bit memory devices for neuromorphic computing [6].

The recent discovery of Sc-doped AlN as a ferroelectric presents a novel and promising avenue for the realization of practical FTJs [7-8]. AlN alloyed with Sc shows large coercive fields, $E_c$, of 2-4.5 MV/cm, which enables scaling to thinner ferroelectric layers, while maintaining a large memory window. When combined with high remnant polarizations – $P_r$, of 80-115 μC/cm$^2$ – this leads to significant resistive switching, due to the strong tunnel barrier modulation, and thus a high On/Off ratio. The more recently reported ferroelectric switching in sub-20 nm Sc-doped AlN at low deposition temperature ($\leq$ 350 °C), allows for these devices to be integrated directly in a CMOS, BEOL-compatible process [9].

Here, we demonstrate $Al_{0.64}Sc_{0.36}N$-based FTJ memristors that are fabricated in a fully BEOL, CMOS-compatible process on a 4-inch Si chip. With 20-nm-thick $Al_{0.64}Sc_{0.36}N$ as a ferroelectric layer with a thin native oxide tunnel barrier layer, the resulting memristors exhibit high performance, with a large self-rectifying ratio > $10^5$, a high On/Off ratio of over 50,000, a stable programmed state over DC cycling, and a retention time longer than 1,000 s at 300K. These results hold promise for future high-performance, post-CMOS compatible NVM.

Fig. 1(a) shows the schematic diagram of a Pt/oxide/ $Al_{0.64}Sc_{0.36}N$/Pt FTJ memristor. A Pt (100 nm) bottom electrode (BE) was deposited by sputtering onto the Si substrate. Next, a 20-nm $Al_{0.64}Sc_{0.36}N$ film was co-sputtered from two separate 4-inch Al (1000 W) and Sc (655 W) targets in an Evatec CLUSTERLINE® 200 II pulsed DC Physical Vapor Deposition System. The deposition was done at 350 °C with $N_2$ gas flow of 20 sccm respectively. Subsequently, top electrode (TE) regions were then patterned using a standard photolithography, as shown in Fig. 1(c), followed by evaporation of a Pt (100 nm) metal top electrode and lift-off process. Fig. 1(b) shows a cross-sectional transmission electron microscopy (TEM) image of the $Al_{0.64}Sc_{0.36}N$ film on Pt BE. In Fig. 1(b) inset, the high-resolution TEM image of the regions enclosed by the red boxes illustrates textured epitaxial growth of the $Al_{0.64}Sc_{0.36}N$ film and a ~4 nm native oxide layer at the ferroelectric surface due to exposure to ambient air.

Because the leakage in our FTJ device is both polarization-dependent and asymmetric in nature, the ferroelectric polarization upon switching was confounded by leakage. To overcome this issue, positive up, negative down (PUND) measurements were performed at low temperature (120 K) and at a relatively high frequency (10 kHz) to suppress leakage and clarify the FE properties of the $Al_{0.64}Sc_{0.36}N$. More details of polarization-electric (P-E) hysteresis measurements at room temperature can be found in our previous work [9]. Fig. 1(c) presents a typical P-E hysteresis loop of the $Al_{0.64}Sc_{0.36}N$ device extracted from the PUND measurements. The measurement indicates a coercive field of 6.5 MV/cm and a remnant polarization of 25 μC/cm$^2$. The coercive field is observed to be slightly larger than the values observed in our DC measurements and the values reported in Ref. [7-9] at 300 K. We posit that this is because the coercive field is reported to significantly increase as temperature drops [10-11] and frequency increases [7, 12-13]. We attribute the relatively lower remnant polarization to the partial ferroelectric switching during the PUND measurement at low temperature. Further, the C-V curve of our device measured at 1 MHz at room temperature (Fig. 1(d)) illustrates a butterfly-shaped loop suggesting a non-linear capacitor that has decreasing capacitance with increasing applied voltage, indicating ferroelectric polarization switching. Similar C-V curves have been reported for FTJ devices based on other FE materials [19-20].

Fig. 2 shows the I–V characteristics of the $Al_{0.64}Sc_{0.36}N$ FTJs in semi-log and linear scale. The blue plots represent non-linear diode-like I-V curves in which polarization points to TE after applying a negative program voltage, whereas the orange plots show I-V curves in which polarization points to BE after applying a positive program voltage. After being programmed by a positive voltage, the resistance changes from high to low, and the polarity of the memristor changes from a negative-forward diode (blue lines) to a positive-forward diode (orange lines). Similarly, it can be observed in negative voltage sweep that the polarity of the memristor changes from that of a positive forward diode to that of a negative-forward diode. Because of the existence of a Schottky barrier between the $Al_{0.64}Sc_{0.36}N$ and the metal electrode (Pt), the device shows self-selective behavior as a diode with > $10^5$ rectification ratio for 5 V, which suppresses sneak currents without additional access transistors or selectors [6]. A large On/Off ratio of 50,000 is obtained in this memristor at a readout bias voltage ~2.5V, which is at least two orders of magnitude above the On/Off ratio previously reported in hafnia-based FTJs [4-5]. A summary of FE memristor characteristics from the literature is presented in Table I, with a focus on the CMOS compatibility, On/Off ratio, and the thickness of the ferroelectric tunnel barrier. It is worth noting that our reported $Al_{0.64}Sc_{0.36}N$ -based FTJ memristor is the only one that has a high ON/OFF ratio while currently being compatible with CMOS BEOL processing.

Furthermore, we studied the effect of polarization on the band diagram and electronic transport in these $Al_{0.64}Sc_{0.36}N$ FTJs. We fitted the forward current through the diode versus applied voltage with well-known tunneling models such as the Fowler–Nordheim (F-N) and Poole-Frenkel (P-F) models since the contribution of trap-assisted tunneling (TAT) could become large in FTJs with a 20-nm-thick ferroelectric barrier [4]. Polarization charge effects induce an asymmetric modulation of the electronic band diagram. As shown in Fig. 3, when *P* is reversed, the steepness of the electronic band diagram is changed, depending on whether the direction of *P* is identical or opposite to the applied electrical field. As show in Fig. 3, when a positive voltage is applied to the TE, the barrier height is – on average– higher when *P* points to the TE than when *P* points to the BE. Electronic band-diagrams of high resistance states (HRS) and low resistance states (LRS) of the FTJ at positive bias, with two tunneling mechanisms, have been sketched in Fig. 3(a) and (b). Among these two transport models, the current-voltage data fits the P-F tunneling model best, as shown Fig. 3(c). The extracted dielectric constant of ~16 of the ferroelectric is close to independent capacitance measurements of 14~15, which is also similar with the values reported by Akiyama *et al*. [21]. For the LRS, the applied electrical field follows the ferroelectric polarization direction. The injected electrons tunnel from highly occupied traps to empty traps and consequently, the current is high. Conversely, if the applied electrical field is opposite to the ferroelectric polarization direction, the electron hopping rate is significantly reduced, as there are fewer occupied electrons. This leads to empty traps and a lower current through the diode.

Preliminary reliability tests have also been conducted. Fig. 4(a) presents data from 10 manually performed DC cycles. Cyclic I-V curves from the same device indicate that the current-voltage characteristics are stable and repeatable. Furthermore, the two different polarization states of the ferroelectric and hence resistance (current) states of the memristor can be programmed as two non-volatile memory states. Readouts at various delay times were carried out to determine retention (Fig. 4(b)). The low and high current/resistance states can be retained for at least 1,000 secs at room temperature without obvious degradation.

In summary, we have demonstrated $Al_{0.64}Sc_{0.36}N$-based FJT memristors that are fabricated in a fully BEOL, CMOS- compatible process on 4-inch Si wafers. With 20-nm-thick $Al_{0.64}Sc_{0.36}N$ as a ferroelectric layer, the resulting memristors exhibit high performance with a large self-rectifying ratio > $10^5$, a high on/off ratio of over 50,000, a stable programmed state over DC cycling, and a retention time longer than 1,000 s at 300K. These results demonstrate that this system has significant potential as a future high-performance post-CMOS compatible non-volatile memory technology.

**Data Availability Statement:** The data that support the findings of this study are available within the article. Additional information and data are available from the corresponding author upon reasonable request.

This work was supported by the DARPA TUFEN program. The work was carried out in part at the Singh Center for Nanotechnology at the University of Pennsylvania which is supported by the National Science Foundation (NSF) National Nanotechnology Coordinated Infrastructure Program (NSF grant NNCI-1542153). The authors acknowledge use of facilities supported by NSF through the Penn Materials Research Science and Engineering Center (MRSEC) (DMR-1720530). TEM sample preparation was performed by Kim Kisslinger at the Center for Functional Nanomaterials, Brookhaven National Laboratory, which is a U.S. DOE Office of Science Facility, under Contract No. DE-SC0012704.

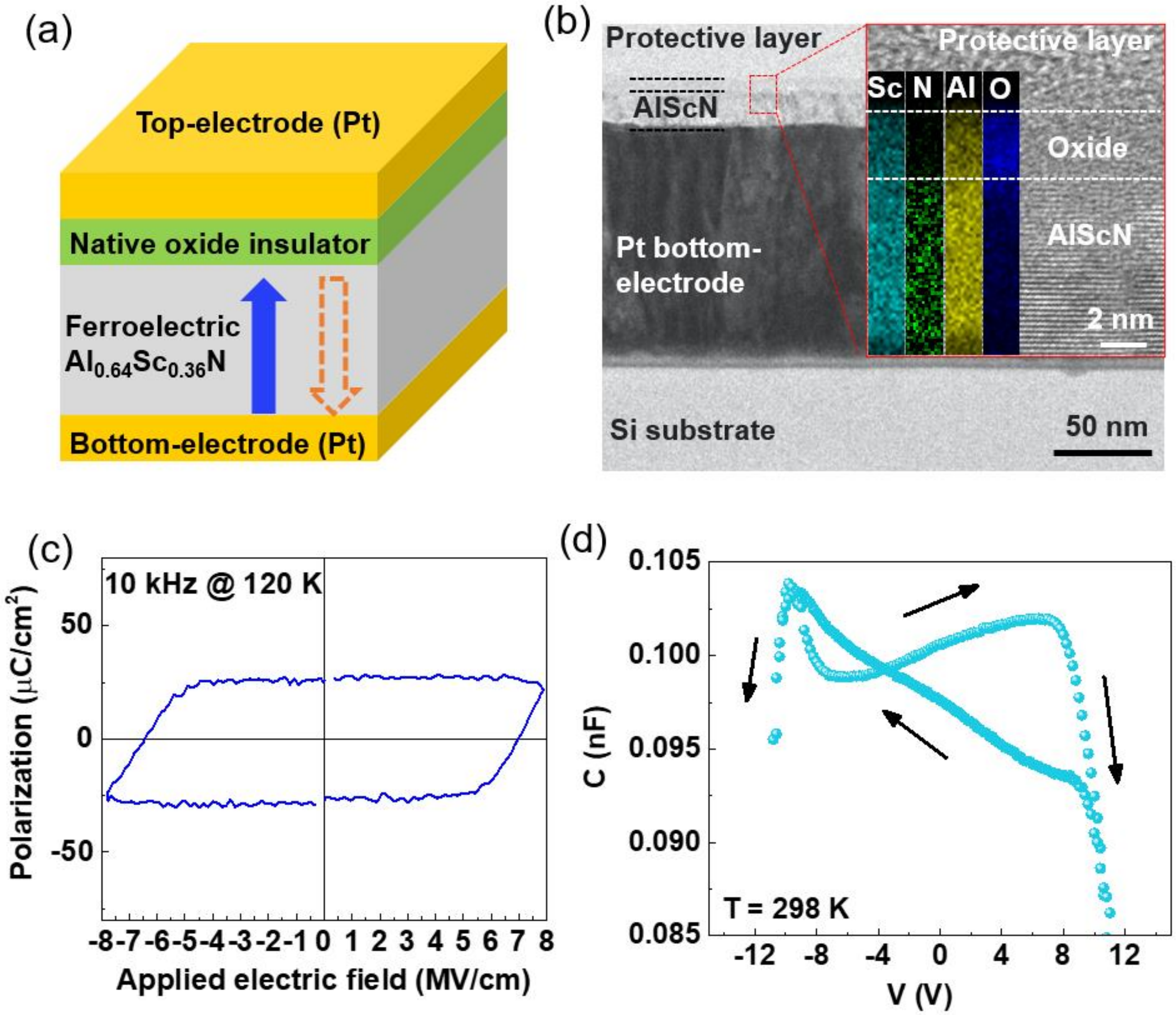

FIG. 1. (a) Schematic of the metal/oxide/$Al_{0.64}Sc_{0.36}N$/metal memristor device. (b) Cross-section transmission electron micrograph of the device showing a Pt bottom electrode grown on the Si wafer substrate, as well as the $Al_{0.64}Sc_{0.36}N$ ferroelectric. Inset, the high-resolution TEM images of the regions enclosed by the red boxes. (c) $Al_{0.64}Sc_{0.36}N$ PUND measurement results measured at 120 K. (d) Representative C-V curve of our devices measured at 1 MHz at room temperature, illustrates a typical butterfly loop.

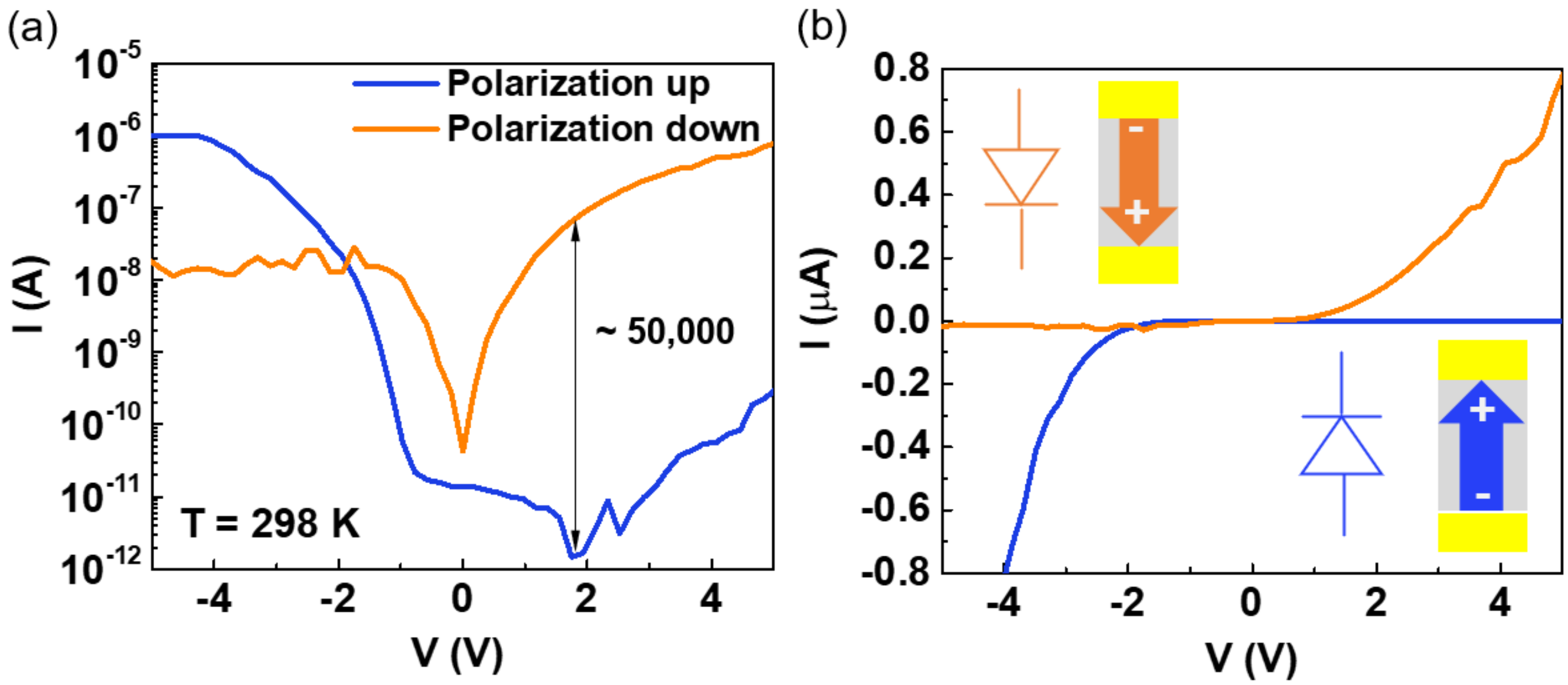

FIG. 2. (a) Semi-log and (b) Linear current-voltage characteristics of the ferroelectric memristor. The blue plots represent nonlinear I–V curve after applying a negative program voltage, whereas the orange plots show I-V sweeps in which the resistance state has been programmed by a positive voltage.

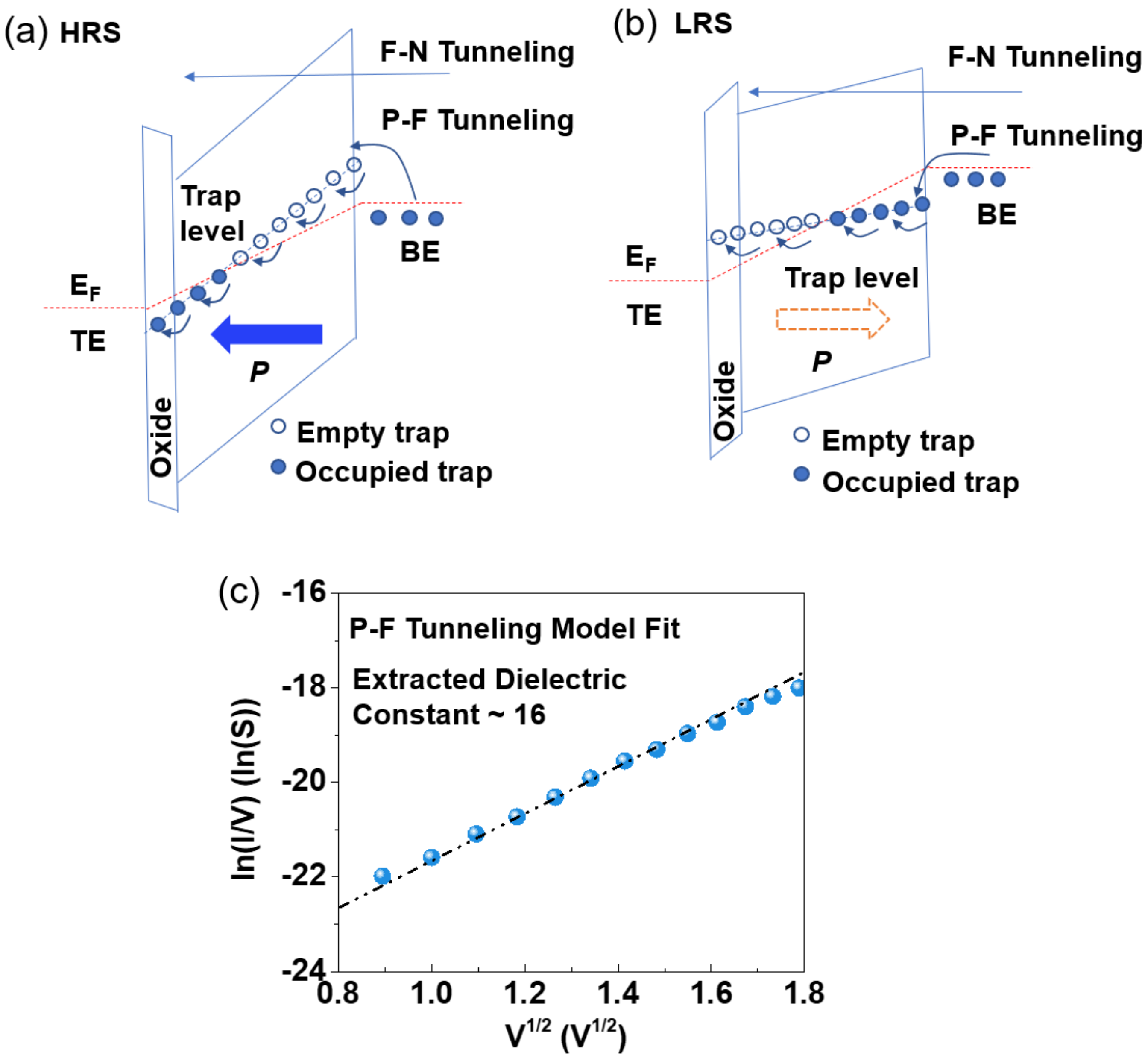

FIG. 3. (a) Electronic band-of HRS and (b) LRS. (c) Fitting of experimental current-voltage data to the Poole-Frenkel tunneling model showing a good fit with extracted dielectric constant of ~16 of the insulator which matches with independent capacitance measurements.

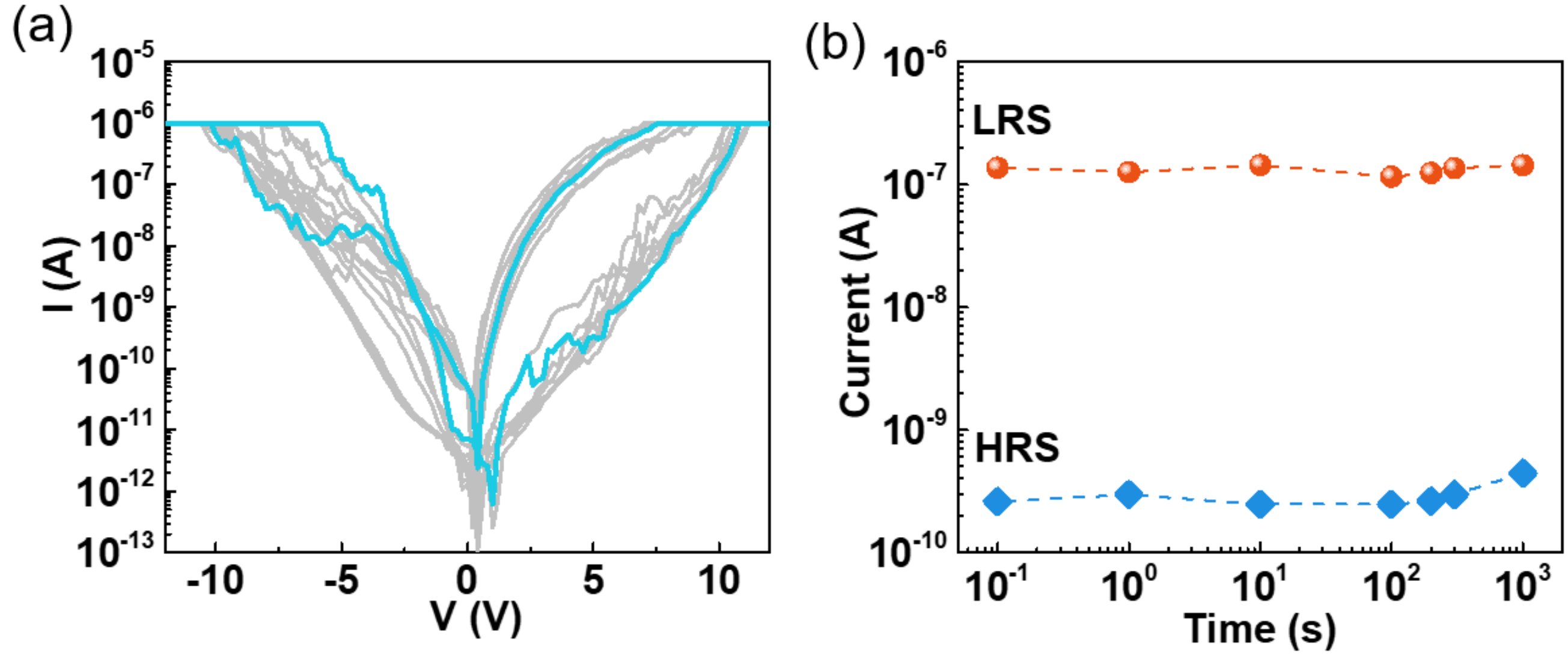

FIG. 4. (a) Cyclic IV curves over manual DC switching cycles showing stability and repeatability of FTJs. (b) Retention of the low and high current/resistance states ~1,000 s, by the readout at 5 V.

TABLE I. Summary of reported ferroelectric memristors.

| **Ferroelectric** | **CMOS Compatibility** | **On/Off Ratio** | **Thickness (nm)** |
|---|---|---|---|
| BFO [14] | Low | 20 | 90,000 |
| BFO [15] | Low | 20,000 | 3 |
| $PbTiO_3$ [16] | Low | 2 | 200 |
| PZT [17] | Low | 300 | 30 |
| BTO [18] | Low | 12,000 | 3.2 |
| BTO [18] | Low | $6 \times 10^6$ | 1.6 |
| Doped $HfO_2$ [4-5] | Medium | 10 | 10 |
| **This work** | High | 50,000 | 20 |